\documentclass{emulateapj}
\usepackage{graphicx}
\usepackage{epsfig}
\usepackage{amsmath}
\usepackage{url}
\begin{document}
\title{Chandra observations of the elusive pulsar wind nebula around PSR B0656+14}
\author{L. B\^{\i}rzan}
\affil{Hamburger Sternwarte, Universit$\ddot{a}$t Hamburg, Gojenbergsweg 112, 21029, Hamburg, Germany; lbirzan@hs.uni-hamburg.de}
\author{G. G. Pavlov}
\affil{Department of Astronomy and Astrophysics, Pennsylvania State University, 525 Davey Lab., University Park, PA 16802, USA}
\author{O. Kargaltsev}
\affil{Department of Physics, George Washington University, Washington, DC, 20052, USA}

\slugcomment{Accepted by ApJ}

\begin{abstract}

PSR B0656+14 is a middle-aged pulsar with a characteristic age $\tau_c=110$ kyr  and spin-down power
$\dot{E}= 3.8\times 10^{34}$ erg s$^{-1}$.
 Using \emph{Chandra} data, we searched for  a pulsar wind nebula (PWN)
 and found evidence of extended emission in a  $3\farcs5$--15$\arcsec$ annulus around the pulsar, with
 a luminosity $L_{\rm 0.5-8\,keV}^{\rm ext} \sim 8\times 10^{28}$ erg s$^{-1}$ (at the distance of 288 pc), which is a fraction of
$\sim 0.05$ of the non-thermal pulsar luminosity. If the extended emission is mostly due to a PWN, its X-ray efficiency,
$\eta_{\rm pwn} = L_{\rm 0.5-8\,keV}^{\rm ext}/\dot{E} \sim
2\times 10^{-6}$, is lower than those of most other known PWNe but similar to that of the
middle-aged Geminga pulsar. The small radial extent and nearly round shape of the putative PWN
can be explained if the pulsar is receding (or approaching) in the direction close to the line of sight.
The very soft spectrum of the extended emission ($\Gamma\sim 8$),
much softer than those of typical PWNe, could be
explained by a contribution from a faint dust scattering halo, which
may dominate in the outer part of the extended emission.

\end{abstract}


\section{Introduction}
Pulsars are rotating neutron stars (NSs) formed in supernova (SN) explosions \citep{gold68}. The diverse NS zoo
 \citep{kaspi10}  includes rotation-powered pulsars (RPPs), magnetars (anomalous X-ray pulsars [AXPs] and soft gamma repeaters [SGRs]), compact central objects (CCOs), thermally emitting isolated neutron stars (TEINSs), and rotating radio transients (RRATs).
The RPPs (hereafter pulsars) are identified via nonthermal pulsed emission, which can be seen from radio
to $\gamma$-rays. They are known to produce relativistic winds which carry a significant fraction of the
rotational energy loss rate \citep{gaen06,karg08,karg10}.  The interaction between the pulsar winds and the ambient medium, either
inside or outside the supernova  remnant (SNR), results in a pulsar wind nebula (PWN), which emits  synchrothron radiation from the radio to GeV
$\gamma$-rays \citep{gaen06}. These PWNe can also be seen in TeV $\gamma$-rays produced via the inverse Compton (IC) scattering of the background photons off the same electrons that emit the synchrotron radiation \citep{karg10,karg12}.

The high angular resolution of the \emph{Chandra} X-ray Observatory has made it possible to image about
70 PWNe at X-ray wavelengths \citep{li08,karg08,karg10,karg13}.
These images reveal complex PWN morphologies,
which depend on the ratio of the relative pulsar's velocity to the speed of sound in the ambient medium, and the parameters of the pulsar and the ambient medium. Young pulsars, which have not left their host SNRs, usually
move subsonically through the hot SNR matter, and their PWNe show torus-jet morphologies \citep[e.g., the Crab pulsar;][]{weis00}.
Older pulsars move supersonically in the cold
interstellar medium, and the shocked pulsar wind is confined by a bow shock ahead of the
moving pulsar, while behind the pulsar, a tail is formed, collimated by the ram
pressure and internal magnetic field \citep[e.g., PSR J1509--5850;][]{karg08b}.

Since the pulsar spin-down power $\dot{E}$ decreases with pulsar's age,
the PWN luminosity, crudely proportional to $\dot{E}$, also decreases, which means that
only a handful of PWNe produced by nearby pulsars is observable with current instruments.
However, such PWNe are particularly interesting because they
 show some unexpected properties, such as the three
tails in the PWN of the 340~kyr old Geminga pulsar \citep{pavl10}.
To better understand the properties of relatively old PWNe, we analyzed archival {\sl Chandra} observations of the middle-aged pulsar B0656+14.

PSR B0656+14, was discovered as a radio pulsar in the second Molonglo pulsar survey \citep{manc78}. It has a period $P=0.385$ s, a period derivative $\dot{P}=5.5 \times 10^{-14}$, a characteristic age $\tau_{\rm{c}}=P/2\dot{P}=110$~kyr,   a spin-down power $\dot{E}=3.8 \times 10^{34}$ erg s$^{-1}$ and a dipole surface magnetic field of $B=
 4.7\times 10^{12}$~G \citep{manc05}. The parallax of the pulsar, determined with Very Long Baseline Array observations, yields a distance $d=288_{-27}^{+33}$~pc \citep{bris03}. Its proper motion, $\mu_\alpha\,\cos\delta = 44.07\pm 0.63$~mas~yr$^{-1}$,
$\mu_\delta = -2.40\pm 0.29$ mas~yr$^{-1}$, corresponds to transverse velocity
$v_\perp\approx 60^{+7}_{-6}$~km~s$^{-1}$ directed eastward.  Furthermore, it is believed that the pulsar is associated with the Monogem ring, a bright diffuse $25^{\circ}$ diameter SNR \citep{thor03}, which is visible in the soft X-ray data from the {\sl ROSAT} All-Sky Survey \citep{pluc96}.

PSR B0656+14 has been extensively studied across the entire electromagnetic spectrum
(e.g., \citeauthor{dura11} \citeyear{dura11} and references therein).
In the radio \citet{welt06} found that, in addition to the
persistent emission, PSR B0656+14 also exhibits very bright radio bursts, similar to the ones found in RRATs \citep{McLa06}. In X-rays, PSR B0656+14
was studied with the \emph{Einstein X-ray Observatory}  \citep{cord89}, \emph{ROSAT}  \citep{finl92,ande93,poss96}, {\sl ASCA} \citep{grei96,kawa96}, {\sl RXTE} \citep{chan99}, {\sl BeppoSAX} \citep{mine02,beck99},  \emph{Chandra} \citep{mars02,pavl02} and \emph{XMM-Newton} \citep{zavl04,DeLu05}.
It was found that the X-ray emission is a combination of thermal and non-thermal components, dominating at lower and
higher X-ray energies, respectively. The non-thermal component, characterized by a power-law (PL) spectrum, is believed to be emitted by relativistic particles in the pulsar magnetosphere. The thermal component can be radiated  from the bulk of NS surface (the thermal soft component) and  from the pulsar polar caps (the thermal hard component). The thermal X-ray emission from PSR B0656+14 was modeled as a sum of two blackbody (BB) components with different temperatures and emitting areas \citep{poss96,grei96,mine02,pavl02,zavl04,DeLu05}.

Multiwavelength spectroscopy of PSR B0656+14 \citep{dura11} has revealed that
extrapolation of the best-fit X-ray PL model to lower and higher photon energies approximately
matches (within a factor of two) the near-infrared/optical and GeV $\gamma$-ray spectra.
This suggests that the optical emission may be synchrotron emission
produced by the same PL electron spectral energy distribution (SED)
 within the same region of pulsar magnetosphere \citep[see also ][]{welt10}.
Additionally, \citet{dura11} found that the optical spectrum
shows evidence either for absorption lines or a cool thermal component from a circumstellar dust disk.

Attempts to find extended X-ray emission around PSR B0656+14 date back to observations with  \emph{Einstein X-ray Observatory} \citep{cord89}. Unsuccessful searches for large-scale extended emission were conducted by  \citet{kawa96} ({\sl ASCA}), \citet{ande93} ({\sl ROSAT}) and \citet{beck99} (combined {\sl ASCA}, {\sl ROSAT} and {\sl BeppoSAX}), but, because of relatively low angular resolution, these studies could not rule out the presence of extended emission on small angular scales. Hints of extended emission on arcsecond scales were found in \emph{Chandra} data \citep{mars02, pavl02} and optical data \citep{shib06}, but those results were inconclusive.

In this paper we present further evidence for the presence of extended emission around PSR B0656+14 and investigate scenarios that might explain its origin and properties.

\begin{deluxetable*}{lccccccccc}
\tablewidth{0pt}
\tabletypesize{\scriptsize}
\tablecolumns{9}
\tablecaption{Observation details \label{counts}}
\tablehead{ \colhead{Instrument}&\colhead{Obs. ID}&\colhead{Date}&\colhead{Exposure\tablenotemark{a}}&\multicolumn{2}{c}{Inner region counts\tablenotemark{b}} &\multicolumn{3}{c}{Extended region counts\tablenotemark{d}}\\
\cline{5-6} \cline{7-9}
\colhead{(Mode)}&\colhead{}&\colhead{}&\colhead{(ks)}&\colhead{Total\tablenotemark{e}}&\colhead{Bgd\tablenotemark{f}}&\colhead{Total}&\colhead{Bgd}&\colhead{${\rm Bgd}_{\rm{PSF}}$\tablenotemark{g}}}
\startdata
ACIS-S (TE) & 2801 & 2001 December 15 & 4.413 & 722 $\pm$ 26 & 2.0 $\pm$ 1.4& 132 $\pm$ 12\tablenotemark{h} & 9.0 $\pm$ 2.8 & 74 $\pm$ 5 \\
ACIS-S (CC) & 2800 & 2001 December 15 & 25.017 & 36420 $\pm$ 215  & 225 $\pm$ 15 &  1257 $\pm$ 35 & 754 $\pm$ 27 & 266 $\pm$ 7   \\
HRC-I   & 3903 & 2003 January 30 &  40.017 & 52290 $\pm$ 230\tablenotemark{c} & 62 $\pm$ 8 & 1233 $\pm$ 35 & 240 $\pm$ 15 & 573 $\pm$ 24
\enddata
\tablenotetext{a}{Total live time on source after cleaning for particle background flares.}
\tablenotetext{b}{Number of counts in the 0.3--8 keV band in the $r<7$ \arcsec circle centered on the pulsar.}
\tablenotetext{c}{In the the $r<3\farcs5$  circle centered on the pulsar there are $51744 \pm 230$ counts, and $15 \pm 4$ counts in the background region.}
\tablenotetext{d}{Number of counts in 0.3--8 keV band in the $3\farcs5 < r < 15''$ annulus centered on the pulsar.}
\tablenotetext{e}{Total source counts.}
\tablenotetext{f}{Counts from a source-free background  region with the same area as the one used to obtain the  total counts for the source, offset $\sim 1'$--$2'$ from the pulsar.}
\tablenotetext{g}{PSF  contribution calculated from a simulated image}
\tablenotetext{h}{$5\pm1$ counts in this region are due to the read-out streak.}
\end{deluxetable*}

 \section{Observations and Data Reduction}

PSR B0656+14 was observed by \emph{Chandra} on 2001 December 15 with the ACIS-S (S3 chip) in Continuous Clocking (CC) and Timed Exposure (TE) modes in `very faint' telemetry format for 25.12 ks and 4.96 ks, respectively (ObsID 2800 and 2801), and on 2003 January 30 with HRC-I (ObsID 3903) for 40.24 ks\footnote{There is also a 40 ks LETG/HRC-S observation from November 1999 (ObsID 130), but the image from this observation is contaminated by diffraction spikes. As a result, we did not use it for the study of the PWN; however, these grating data were used with success to study the pulsar in detail \citep{mars02}.}. The TE mode observation was done with the default frame time of 3.2~s  (see Table \ref{counts}).

We reduced the data using CIAO 4.0.1 (CALDB version 3.4.2). The ACIS data were filtered for the standard good grades.
To achieve the sharpest resolution, pixel randomization was removed from the ACIS images, and subpixelization \citep{tsun01,mori01}
was performed. The TE-mode data in the core of the  point spread function (PSF) were affected by a significant pileup (the pileup fraction is $\approx 65\%$). Therefore, we made spectral fits of the pulsar emission  to the CC-mode data, which are unaffected by pileup thanks to the much shorter frame time.

We extracted the spectra from the ACIS data using the \emph{CIAO} tool \emph{psextract}\footnote{See \url{http://cxc.harvard.edu/ciao4.2/ahelp/psextract.html.}}. Before the extraction, the data were cleaned of
flares, resulting in useful integration time of 4.41 ks for the TE-mode data (no strong flares were identified in the CC-mode data). For the TE mode data, a circular region of $\approx 1''$ radius, centered on the pulsar, was removed as  those data were strongly contaminated by pileup. Unless otherwise mentioned, spectral fits were made with XSPEC v.12.3.1 in the 0.3--8 keV band. The input spectra used for the simulator \emph{ChaRT}\footnote{See \url{http://cxc.harvard.edu/chart}} were produced using the \emph{Sherpa} package (see Section \ref{ps_sim}). Background spectra were extracted from source-free regions on the same CCD chip, $\sim 1'$--$2 '$ from the pulsar.

The HRC-I data did not require reprocessing, filtering for good grades or the sub-pixelization treatment. Table \ref{counts} lists the observation details and source counts in various regions for the 3 observations.

\section{Image analysis}\label{S:images}

\begin{figure*}
\includegraphics[scale=0.623,angle=0]{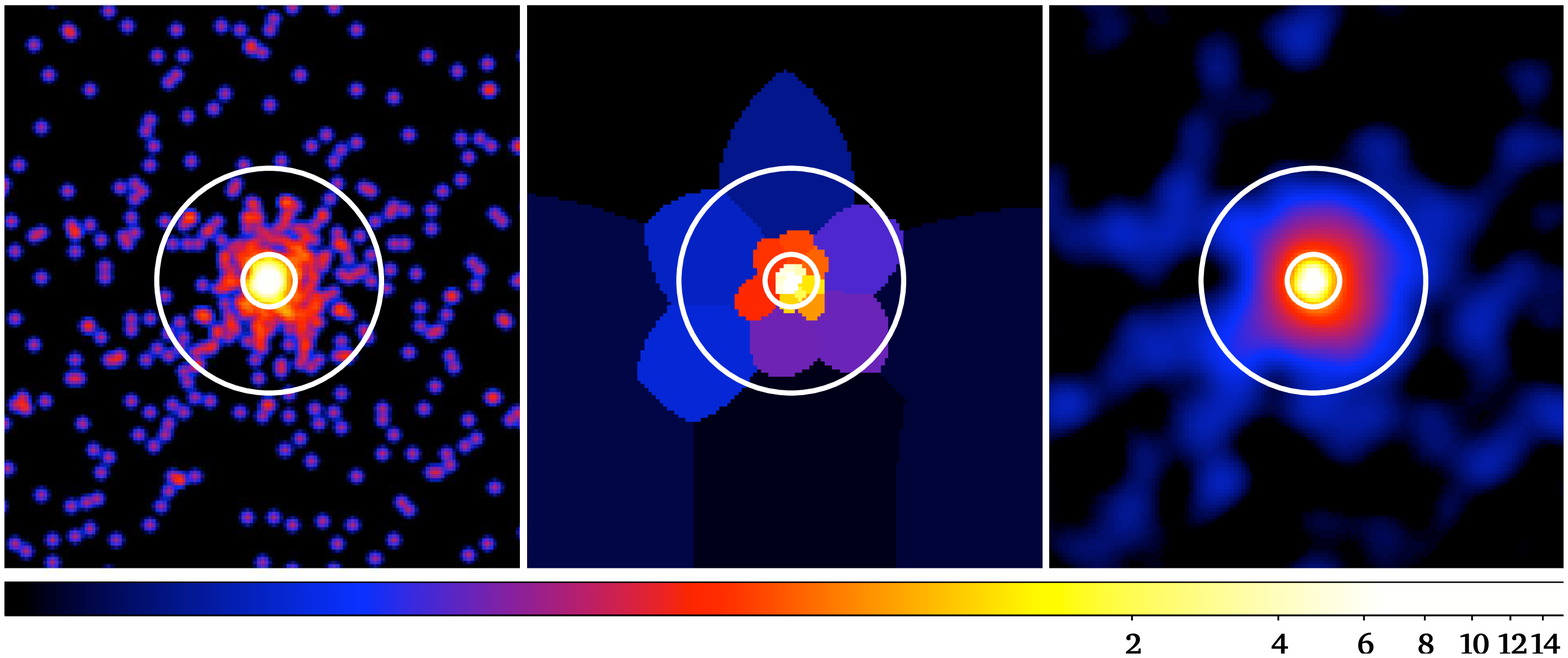}
\includegraphics[scale=0.60,angle=0]{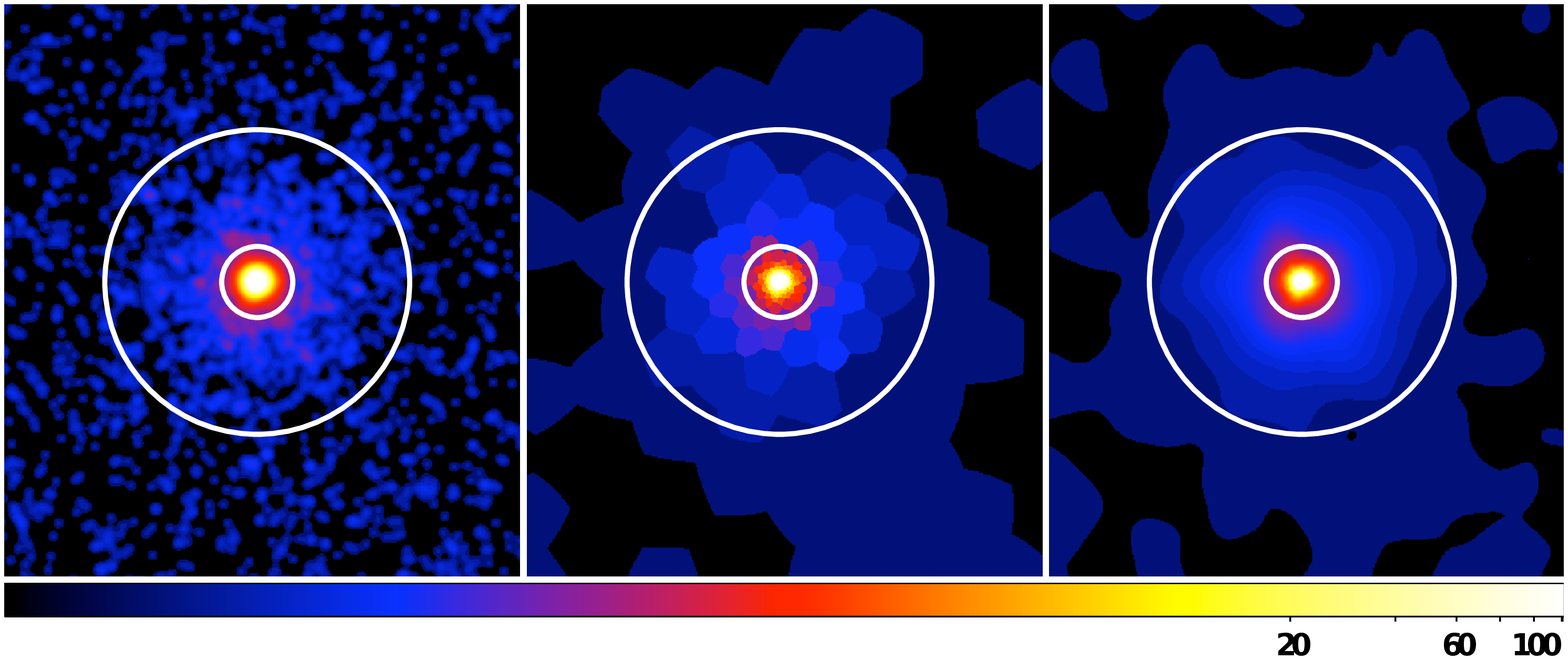}
\caption{Images of PSR B0656+14 from the ACIS-S
(upper panels) and HRC-I
(lower panels) data; the color bars are in units of counts per pixel. \textit{Left}:
The images are smoothed with a simple Gaussian smoothing with 1 ACIS pixel
(0\farcs492)
and 3 HRC pixels (0\farcs395) widths.
 \textit{Middle}: Adaptively binned images with a minimum signal-to-noise per bin of 5.  \textit{Right}: Adaptively smoothed images produced with
 \textit{csmooth}. The circles show radii of $3\farcs5$  and $15''$.
 \label{F:acis_hrc_images}}
\end{figure*}

Smoothed and binned images for the ACIS-S (TE-mode) and HRC-I data are shown in Figure \ref{F:acis_hrc_images}. For a 7\arcsec~diameter region centered on the pulsar, the count rates are about  0.13 and 0.16 counts s$^{-1}$  for the HRC-I and ACIS-S observations, respectively
(see Table~\ref{counts}). The images show a hint of an extended asymmetric structure beyond $r\approx 3\arcsec$.

\subsection{Point Source Simulations}\label{ps_sim}
To further  constrain the presence of the putative extended emission, we simulated point source images for both the ACIS-S and HRC-I observations using ChaRT and MARX\footnote{See \url{ftp://space.mit.edu/pub/cxc/marx/v4.0/marx\_4.0\_manual.pdf}}.
For the simulations we used a larger exposure time of 200 ks (to reduce the statistical errors) and the pulsar spectrum measured from the ACIS-S CC-mode data (see Section \ref{S:spectra} for details of the modeling).

We ran MARX (version 4.2.0) using the output from ChaRT, specifying the starting date,
the pointing right accession, declination, roll angle, the source coordinates and the detector type.
We ran the simulations for three different values of the \textit{ditherblur}
parameter (0\farcs15, 0\farcs25 and 0\farcs35); larger values produce greater blurring. In order to simulate the  ACIS-S TE mode data, we used the pileup model by \citet{davis01} included in MARX, for a few values of the `grade migration' parameter $\alpha$.

\subsection{Surface Brightness Profiles}\label{S:profiles}

\begin{figure*}
\plottwo{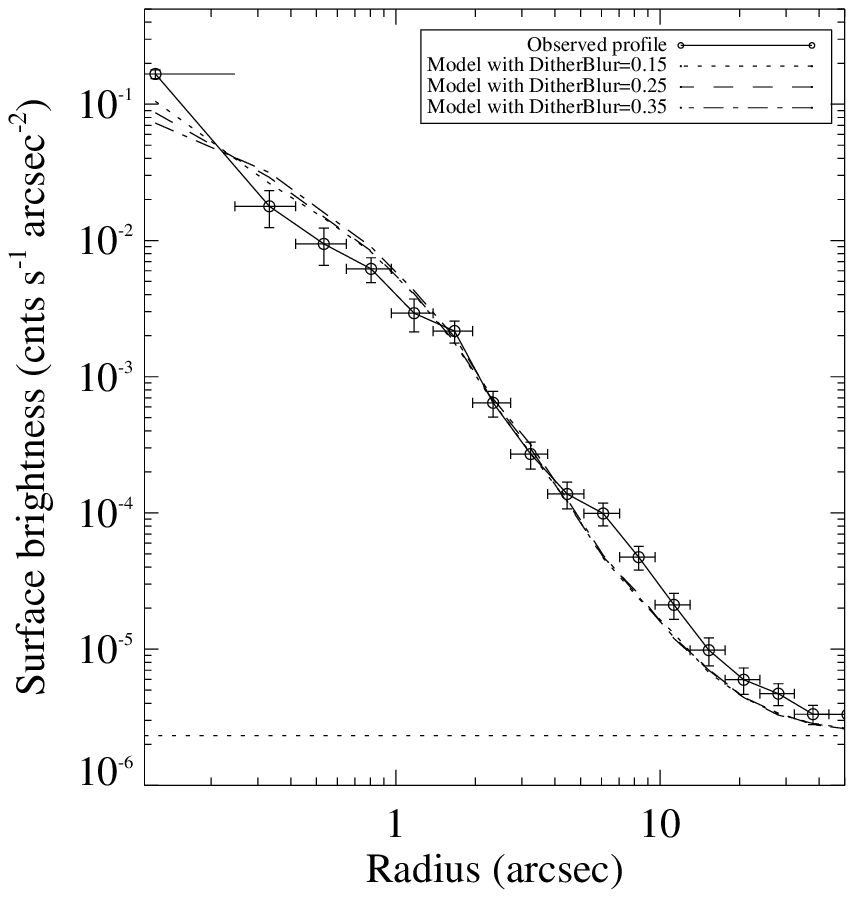}{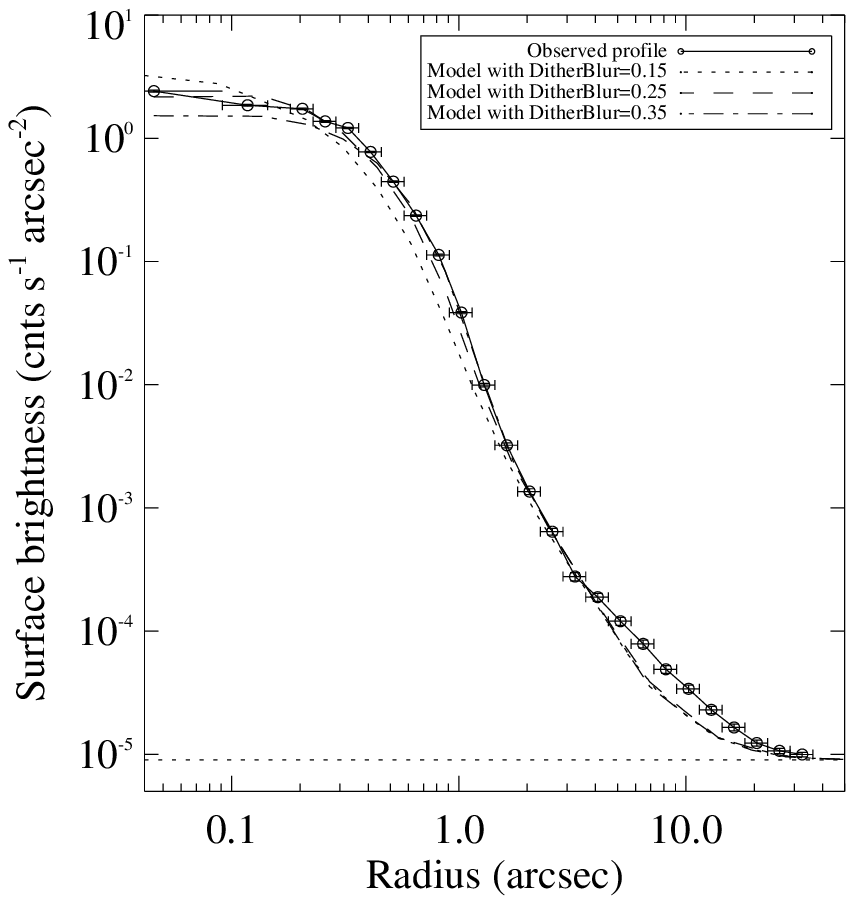}
\caption{Surface brightness profiles in the 0.3--8 keV band for the ACIS-S TE-mode observation (symbols and error bars) on the left, and for the HRC-I observation (symbols and error bars) on the right. The simulations use $\alpha=0.5$ and three different values of the \textit{ditherblur} parameter.
The horizontal dotted lines show the background levels.
\label{F:acis_hrc_sb_dither}}
\end{figure*}

\begin{figure}
\plotone{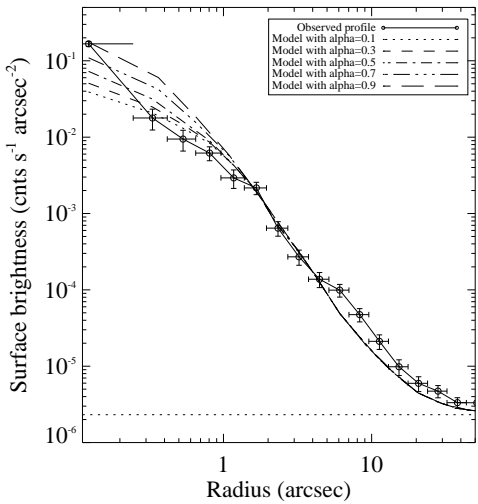}
\caption{Surface brightness profiles in the 0.3--8 keV band for the ACIS-S TE-mode observation (symbols and error bars) and simulations with \textit{ditherblur} = 0\farcs35 and five different values of
$\alpha$.
\label{F:acis_sb_alpha}}
\end{figure}

In Figures \ref{F:acis_hrc_sb_dither} and \ref{F:acis_sb_alpha}, we compare the surface brightness profiles derived from the images
with those from the point source simulations. The ACIS-S profiles have been extracted
in the 0.3--8 keV band. The backgrounds of $(2.23 \pm 0.84) \times 10^{-6}$ and $(9.03\pm 1.14) \times 10^{-6}$~cnts~s$^{-1}$~arcsec$^{-2}$ for the ACIS-S and HRC-I observations, respectively, were derived from source-free regions of the image, approximately $1'$--$2'$
from the pulsar. As one can see from Figure \ref{F:acis_hrc_sb_dither} (left panel), the simulated surface-brightness profiles with different \textit{ditherblur} values are very similar for $r>2\arcsec$.

For the ACIS-S TE-mode observation, the match inside $r\approx 1$ \arcsec\ is poor,
perhaps due to too crude modeling of the pileup effects (undersampling of the PSF in the image may also contribute to some of the uneven appearance of the profile in this region). One important effect of pileup is `grade migration':
it is likely that the grade assigned to piled events `migrates' to a value inconsistent with a real photon (a `bad grade').
Filtering the data to remove bad grades affects the PSF shape. In the simulations, the grade-migration parameter $\alpha$ controls the degree of grade migration. Figure \ref{F:acis_sb_alpha} shows the profiles for a number of different values of $\alpha$, all for a \textit{ditherblur} value of 0\farcs35. While $\alpha$ has a significant effect on the profile within $r\lesssim~1$~\arcsec, no value of $\alpha$ appears to produce a good match to the observed profile in this region. However, as with the \textit{ditherblur} parameter, varying  $\alpha$ has no effect on the profiles beyond the innermost regions.

For the inner few arcseconds of the HRC-I profile (Figure \ref{F:acis_hrc_sb_dither}, right panel), the simulations with \textit{ditherblur}=0\farcs25  and 0\farcs35  seem to give better matches to the observed profiles than the simulation with \textit{ditherblur}~$=0\farcs15$. Since the simulation with \textit{ditherblur}~$=0\farcs25$  works best only for the inner 0\farcs3, and the simulation with \textit{ditherblur}~$=0$\farcs35 works better  between 0\farcs3  and 3\arcsec, we used \textit{ditherblur}~$=0$\farcs35 in the subsequent analysis to ensure  the most conservative estimate of the diffuse emission level.

In both the ACIS-S and HRC-I data, a clear excess is apparent in the range $3\farcs5 < r < 15''$. Between these radii, we find a total of $44 \pm 12$ excess counts in the ACIS-S data (0.3--8 keV) and $420\pm 45$ excess counts in the HRC-I data (see Table \ref{counts}). As we have demonstrated above, the simulated profiles are insensitive to changes in the \textit{ditherblur} or $\alpha$ parameters beyond the innermost regions ($r\lesssim 2\arcsec$). Therefore, this excess is robust  with respect to the changes in the simulation parameters. The fact that the excess appears at the same range of radii for both the ACIS-S and HRC-I observations rules out errors in calibration of the individual detectors. However, we cannot exclude the possibility that the apparent excess might be due to an intrinsic inaccuracy of the PSF model used in the ChaRT and MARX simulators. This inaccuracy must then be present in the HRMA PSF model.  To explore this possibility, we repeated the above analysis using HRC-I data for two \emph{Chandra} calibration sources, 3C273 and AR Lac (see the Appendix for details), and in neither calibrator do we see evidence for excess emission relative to the simulated profile such as that seen in the HRC-I observation of PSR B0656+14 (see Figure \ref{F:3c273_ARLac} in the Appendix). Furthermore, in Figure \ref{F:all_obs} (see the Appendix) we directly compare the surface brightness profiles for the HRC-I observations of PSR B0656+14, 3C273 and AR Lac. It is clear from this comparison that the excess emission seen in the profile of PSR B0656+14 between $3\farcs5 < r < 15''$ is not present in those of the two calibration sources.

\subsection{Model-subtracted Images}

\begin{figure*}
\plotone{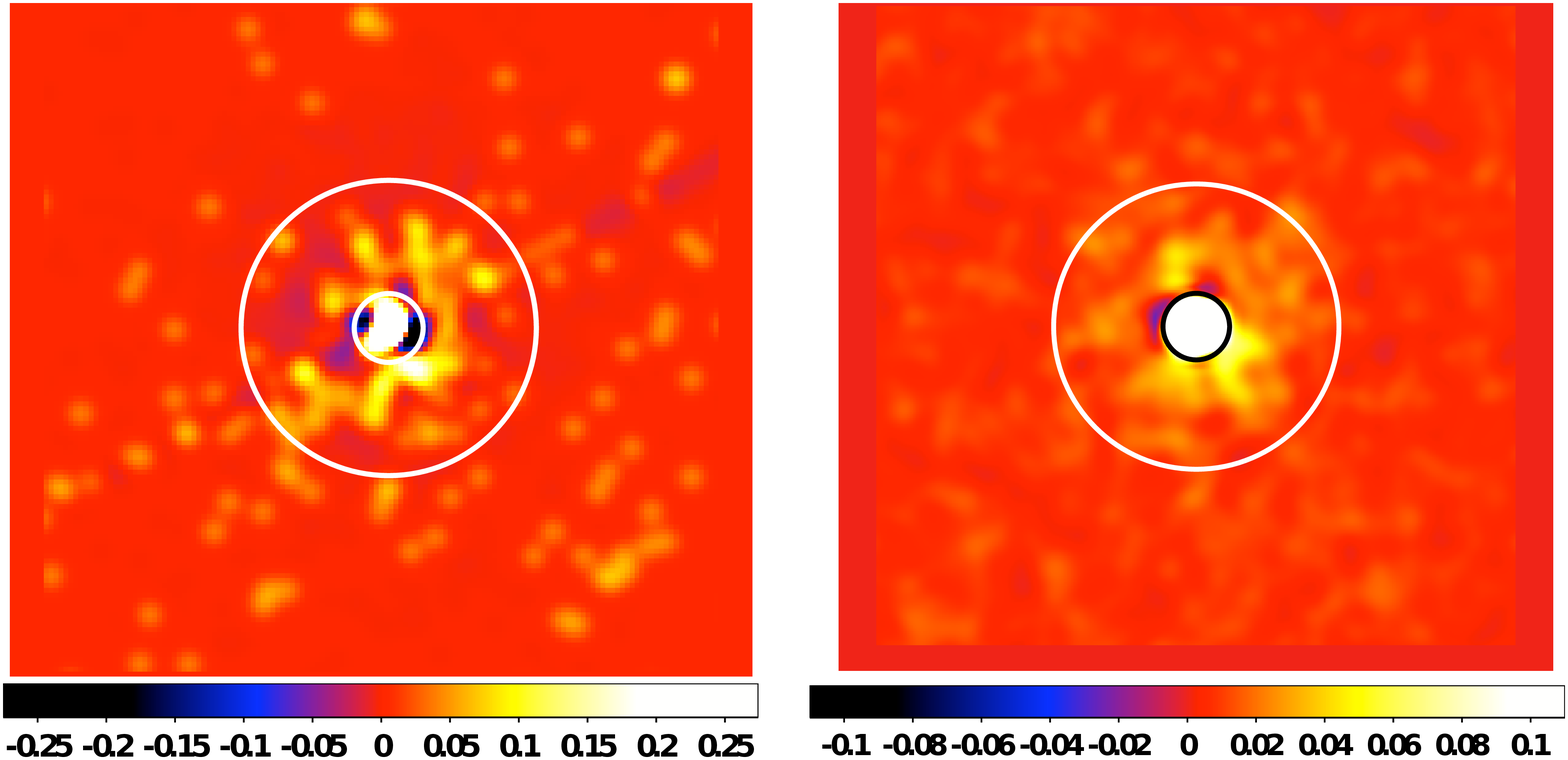}
\caption{\emph{Left:}
ACIS-S image in the 0.3--8 keV band
 after subtracting the simulated point source (see text); the color bars are in units of counts per pixel. The subtracted image is then smoothed with a $r=$2\farcs46 Gaussian kernel.
\emph{Right:} Model-subtracted HRC-I image,
smoothed with a Gaussian kernel of the same size as used in the left panel. The inner and outer circles indicate radii of 3\farcs5
and 15\arcsec, respectively.
 \label{F:images_sub}}
\end{figure*}

To visualize the putative extended emission around the pulsar,
we subtract off a model of the PSF from the observed images.
Figure \ref{F:images_sub} shows the residual
images for the ACIS-S and  HRC-I data produced by subtracting the simulated point source created with \textit{ditherblur} = 0\farcs35 and $\alpha=0.5$. Due to the low number of counts in the residual images, a Gaussian smoothing with a kernel of $r=$2\farcs46 was applied to reduce the noise. In both images, residuals are apparent in the inner few arcsec, in agreement with  the radial profiles presented in Section \ref{S:profiles}. The excess extended emission in the \textbf{$3\farcs5$--15$\arcsec$} annulus is also clearly visible.

\subsection{Asymmetry of the extended emission}\label{asymm_sec}

To examine the asymmetry of the putative extended emission, we measured the numbers of counts in 4 equal quadrants of \textbf{$3\farcs5 < r < 15''$}  annuli
in the ACIS-S and HRC-I images. The quadrants are oriented to the north, east, south, and west, and centered on the pulsar's core. The choice of such quadrants is motivated by the expectation of an asymmetry caused by the proper motion of the pulsar toward East.
In Table \ref{T:counts}, we list the total counts in the four quadrants. A faint readout streak present in the ACIS-S image passes through the east and west quadrants. By comparing the counts in the streak to those from a local background (both situated $\sim 1'$--$2'$
from the pulsar), we estimate the streak's surface brightness (in a region 10\arcsec\ wide) to be $(2.24 \pm 0.65) \times 10^{-2}$~counts~arcsec$^{-2}$ between 0.3--8 keV, thus contributing $\approx 2$--3 counts to each of these two quadrants.

According to Table \ref{T:counts}, the maximum difference in the numbers of counts in different quadrants is
$54\pm 25$ and $13\pm 8$ in the HRC-I and ACIS-S images, respectively. It means
that statistical significance of the apparent asymmetry does not exceed $2.2\sigma$.

\begin{deluxetable}{lccc}
\tablewidth{0pt}
\tablecaption{Counts in four quadrants between 3\farcs5--15\arcsec
  \label{T:counts}}
\tablehead{ \colhead{}&\colhead{ACIS-S counts} &\colhead{HRC-I counts}\\
\colhead{Region}&\colhead{0.3--8 keV} &\colhead{}}
\startdata
North Quadrant & $31\pm6$ & $314\pm18$  \\
East Quadrant  & $25\pm5$ & $284\pm17$ \\
South Quadrant & $38\pm6$ & $338\pm18$ \\
West Quadrant  & $38\pm6$ & $297\pm17$
\enddata

\end{deluxetable}

\vspace{10.0em}

\section{Spectral Modeling}\label{S:spectra}

In this section, we describe the spectral modeling of the pulsar emission and of the spatially extended excess emission. The pulsar emission has been studied extensively previously, and a detailed analysis of this emission is beyond the scope of this paper. However, since the PWN is located close to the pulsar, we need to account for the contribution to the extended emission of the pulsar's PSF wings. In order to model the PSF, we need the pulsar spectrum, which we obtain from the CC-mode data. For discussions and detailed analyses of the pulsar emission see \citet{pavl02}, \citet{mars02}, \citet{zavl04} and \citet{DeLu05}.

\subsection{Pulsar Emission Spectrum}\label{pulsar_mod}

\begin{figure}
\includegraphics[scale=.5,angle=270]{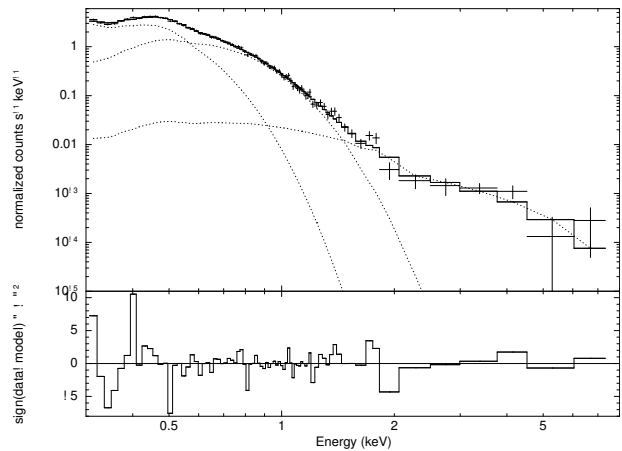}
\caption{The spectrum of PSR B0656+14 within a 1D aperture of 7\arcsec\ width from the CC-mode data (points with error bars). The best-fitting {\tt wabs} $\times$ [BB$_{\rm high}+$ BB$_{\rm low}+$ PL] model is shown by the solid line. The dashed lines show the model components.
Lower panel shows the contribution of the spectral bins to the
$\chi^2$ statistic.
  \label{F:cc_spectrum}}
\end{figure}

We extracted the pulsar spectrum between 0.3 and 8 kev from the CC-mode data
(unaffected by pileup) within a segment of 7\arcsec\
width centered on the source (there are 36420 $\pm$ 215 counts in this region; see Table \ref{counts}). This region was chosen to ensure that most of the pulsar emission was contained within it (e.g., for an ACIS TE observation, a region with a diameter of 7\arcsec\ should include $\approx$ 95\% of counts of a point source with a nearly monochromatic spectrum at 2 keV; due to projection, an even higher fraction should be included in our CC-mode spectrum).  Furthermore, this region was chosen to minimize possible contamination from the extended emission, but some contaminating emission
is still present in the one-dimensional (1D) image due to the projection effect. The background was extracted from a region of the same size from a source-free area $\approx 80$\arcsec\ from the pulsar (containing 225 $\pm$ 15~counts). The source spectrum was binned to a minimum of 30 counts per bin, and fits were made in XSPEC in the 0.3--8 keV band using the three-component model plus galactic absorption \citep[e.g.,][]{grei96}:
{\tt wabs} $\times$ [BB$_{\rm high}+$ BB$_{\rm low}+$ PL], where
{\tt wabs} is the Galactic absorption component which uses \citet{ande82} relative abundances, and BB$_{\rm high}$ and BB$_{\rm low}$ are the BB components with higher and lower temperatures.
The hydrogen column density $N_H$ of the {\tt wabs} component, the BB temperatures, and the photon index $\Gamma$ were allowed to vary. The resulting best-fit parameters (corresponding to the 90\% confidence level) are shown in Table \ref{pulsar_fluxes}, and the resulting fit is shown in Figure \ref{F:cc_spectrum}. The best-fit value for the hydrogen column density, $N_{\rm H}=4.3^{+0.9}_{-0.6} \times 10^{20}$~cm$^{-2}$, is a factor of $\approx 2$ larger than the one found by \citet{mars02}, but agrees with XMM-Newton results from \citet{DeLu05}. The reason for this discrepancy is not clear; it could be due to differences in the fitted energy ranges \citep[we use energies above 0.3~keV;][use energies above 0.15~keV]{mars02}.

To calculate the total absorbed and unabsorbed fluxes, we re-fitted the data by adding a CFLUX component to the model (with XSPEC 12.5.0), with the normalization of the power law component fixed at the previously found value. The resulting total absorbed and unabsorbed fluxes in the 0.5--8.0 keV energy range are listed in Table \ref{pulsar_fluxes}. As the $\chi^2 $ value shows, the above fit is only marginally acceptable, but it is sufficient for our purpose (see Section \ref{model_ext}).

For the distance $d = 288_{-27}^{+33}$ pc \citep{bris03}, the unabsorbed pulsar luminosity in the 0.5--8 keV band
is $L_{\rm psr} = 3.0_{-0.6}^{+0.7}\times 10^{31}$ erg s$^{-1}$, and the unabsorbed luminosity of the PL component in the same energy band is
$L_{\rm psr}^{\rm nonth}=1.5_{-0.4}^{+0.8}\times 10^{30}$ erg s$^{-1}$.

\begin{deluxetable*}{cccccccccc}
\tablewidth{0pt}
\tabletypesize{\scriptsize}
\tablecolumns{10}
\tablecaption{Fits of the PSR B0656+14 spectrum \label{pulsar_fluxes}}
\tablehead{ \colhead{$N_{\rm H,20}$\tablenotemark{a}}&\colhead{$kT_{\rm low}$\tablenotemark{b}}&\colhead{$\mathcal{N}_{\rm low}$\tablenotemark{c}}&\colhead{$kT_{\rm high}$\tablenotemark{d}}&\colhead{$\mathcal{N}_{\rm high}$\tablenotemark{e}}&\colhead{$\Gamma$} &\colhead{$\mathcal{N}_{-5}$\tablenotemark{f}}&\colhead{$\chi_{\nu}^{2}$/dof}&\colhead{$F^{\rm abs}_{-12}$\tablenotemark{g}}&\colhead{$F^{\rm un}_{-12}$\tablenotemark{h}}}
\startdata
 $4.3^{+0.9}_{-0.6}$ & $59^{+3}_{-6}$ & $234^{+144}_{-45}$  & $113 ^{+5}_{-8}$  & $2.57^{+2.18}_{-0.90}$  & $2.30^{+0.68}_{-0.57}$ &  $4.21^{+3.88}_{-1.71}$ & 1.41/78 & $2.35\pm 0.04$ & $3.05 ^{+0.15}_{-0.11}$
\enddata
 \tablecomments{The fits from {\tt wabs} $\times$ [BB$_{\rm low}+$ BB$_{\rm high}+$ PL] model in the 0.3--8 keV energy range from CC data in a segment of 7\arcsec\ width
centered on the source.
}
\tablenotetext{a}{The hydrogen column density of the {\tt wabs} component in units of $10^{20}$ cm$^{-2}$.}
\tablenotetext{b}{The temperature of the
BB$_{\rm low}$ component in units of eV.}
\tablenotetext{c}{The normalization of the
BB$_{\rm low}$ component in units of  (km/288 pc)$^{2}$.}
\tablenotetext{d}{The temperature of the
BB$_{\rm high}$ component in units of eV.}
\tablenotetext{e}{The normalization of the
BB$_{\rm high}$ component in units of  (km/288 pc)$^{2}$.}
\tablenotetext{f}{The normalization of the PL component in units of $10^{-5}$ photons keV$^{-1}$ cm$^{-2}$ s$^{-1}$ at $E=1$ keV.}
\tablenotetext{g}{Total absorbed flux in the 0.5--8 keV energy range, in units of $10^{-12}$ erg s$^{-1}$ cm$^{-2}$.
 }
\tablenotetext{h}{Total unabsorbed flux in the 0.5--8 keV energy range,
 in units of $10^{-12}$ erg s$^{-1}$ cm$^{-2}$. The unabsorbed flux in the 0.5--8.0 keV energy range for only the PL component is $0.15 ^{+0.07}_{-0.03}$ $10^{-12}$ erg s$^{-1}$ cm$^{-2}$, and for the PL plus high energy BB components it is $1.87 ^{+0.27}_{-0.03}$ $10^{-12}$ erg s$^{-1}$ cm$^{-2}$ }
\end{deluxetable*}

\subsection{The Spectrum of the Extended Emission}\label{model_ext}

To constrain the physical properties of the excess emission seen within $3\farcs5 \lesssim r \lesssim 15''$, we used both TE and CC-mode spectra. The TE spectrum was extracted from a circular annulus centered on the pulsar with an inner radius of 3\farcs4 and an outer radius of 14\farcs8. For the CC mode, we extracted the spectrum from two 12\arcsec\ long segments separated by 3\farcs5 from the 1D pulsar image. For the TE-mode spectrum, since the region is extended, we made weighted responses using the CIAO tools \textit{mkwrmf} and \textit{mkwarf}. In the 0.3--8 keV band where fitting was done, there were a total of 132 $\pm$ 12~counts in the TE data and 1257 $\pm$ 35~counts in the CC data (see Table \ref{counts}).

Since source-free regions of the same area ($\sim$ 80\arcsec\ away) contain only about 9 counts for the TE data and 754 $\pm$ 27~counts for the CC data,
the dominant component of the `background' for the extended emission is the emission from the wings of the pulsar's PSF.
To account for this emission, we included the three-component pulsar model discussed in Section \ref{pulsar_mod} in the fits to the excess. Since the PSF is energy dependent, spectra extracted from the PSF wings differ somewhat from those of the full source (i.e., those within the 95\% enclosed-counts radius). To account for the effects of this difference, we re-fit the {\tt wabs} $\times$ [BB + BB + PL] model to the spectrum extracted from the $\approx$~3\farcs5--15\arcsec circular annular region of the MARX simulation (for the TE-mode), and in two 12\arcsec\ wide rectangular regions with the height equal to the size of the chip, located 3\farcs5 away from the pulsar (for the CC-mode).  In the case of CC mode, we extracted the spectra from an ACIS-S TE-mode simulation (using parameters, such as the roll angle, appropriate for the CC mode observation). For the TE case, this model contributes 74 out of 132 total counts in this region. Lastly, $\approx 4$--6~counts in this region are due to the read-out streak in the ACIS-S data. The total net counts due to the putative excess between 0.3--8 keV for the TE mode is therefore $\approx 44 \pm 12$ (see Table \ref{counts}). For the CC mode, there were $266 \pm 7$ counts in the two regions described above (see Table \ref{counts}). As a result, after subtraction of the sky and pulsar backgrounds, there remain 237 $\pm$ 50 counts from the extended emission
in the CC-mode data. 

To model the `background' emission from the pulsar's PSF wings, the simulated pulsar spectra extracted from these regions were fit  with a model consisting of two absorbed BB components. The fit did not require the PL component. The resulting parameter values\footnote{Note that these values are different
from those inferred from the fit of the pulsar spectrum (see Table 3). This BB+BB model should be considered as a convenient empirical characterization of the PSF wings' spectrum, but any other choice of model that adequately represents the spectrum could have been used (such as PL).}  for the TE-mode case  were $kT_{\rm low}=92$~eV, $kT_{\rm high}=178$~eV,  $\mathcal{N}_{\rm low}=0.19$~(km/288~pc)$^{2}$, $\mathcal{N}_{\rm high}=0.0013$~(km/288~pc)$^{2}$, where $N_{\rm{H}}$ was fixed to $N_{\rm H}=4.3 \times 10^{20}$~cm$^{-2}$ (the best fit from modeling the pulsar emission, see Section \ref{pulsar_mod}). The  resulting parameter values for the CC-mode case were $kT_{\rm low}=88$~eV, $kT_{\rm high}=147$~eV,  $\mathcal{N}_{\rm low}=0.125$~(km/288~pc)$^{2}$, $\mathcal{N}_{\rm high}=0.0035$~(km/288~pc)$^{2}$ with fixed $N_{\rm H} = 4.3 \times 10^{20}$~cm$^{-2}$. These parameters were used to define the background pulsar emission included in the fit to the extended emission.

\begin{figure*}
\plottwo{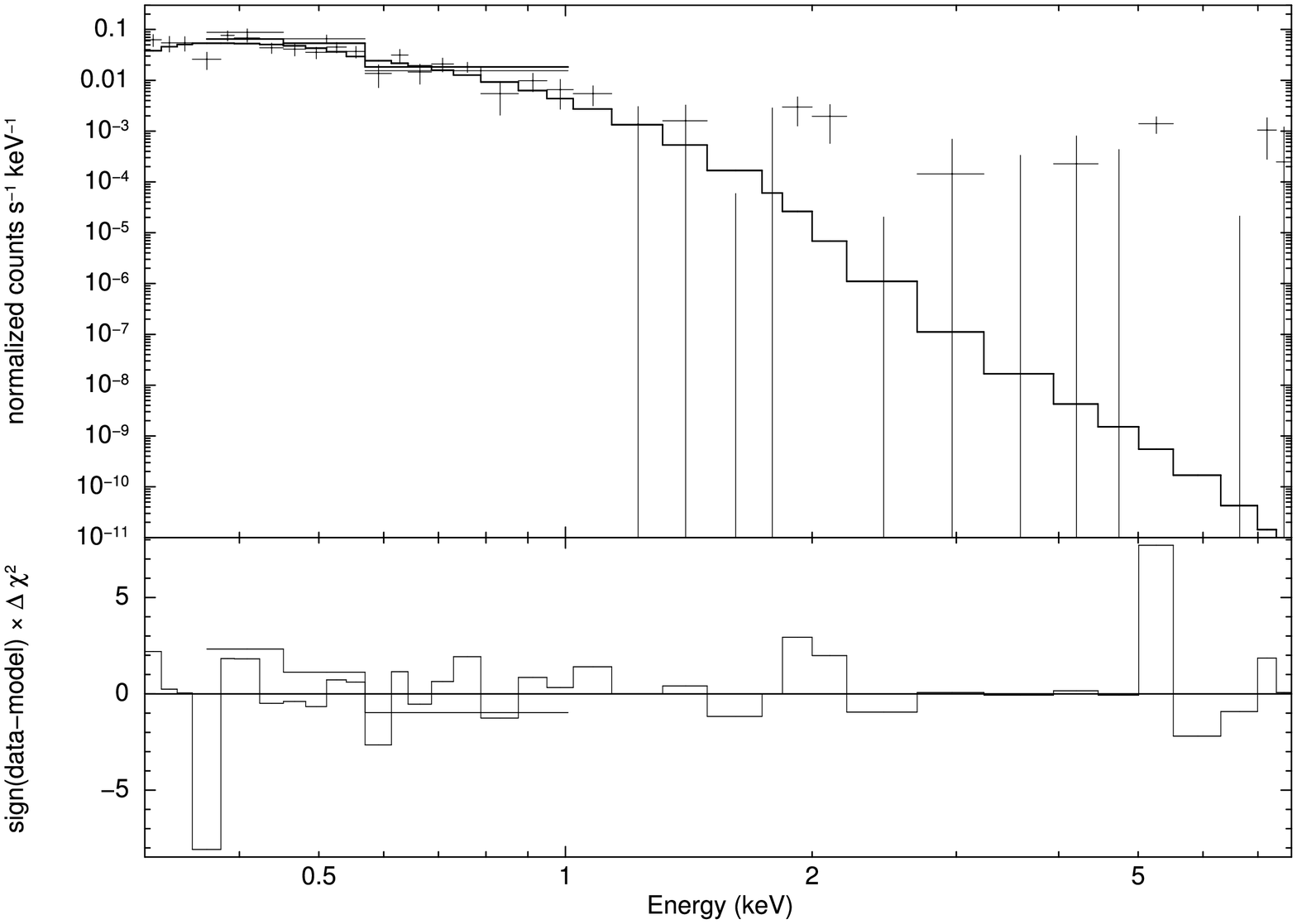}{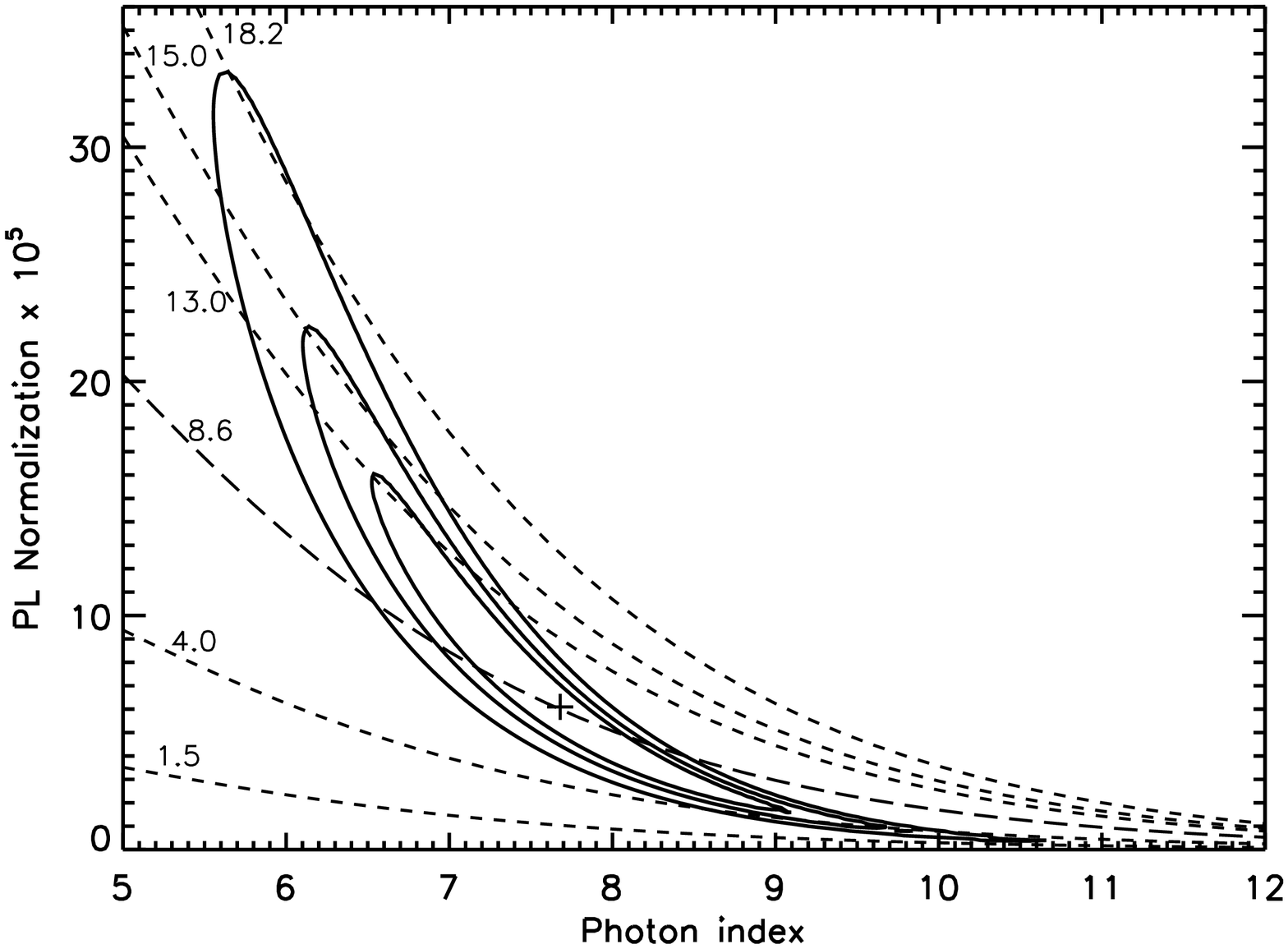}
\caption{\emph{Left}: The data and best-fitting model from the simultaneous fit of both CC- and TE-mode spectra, extracted from the 7--30~pixel ($3\farcs4 < r < 14\farcs8$) annulus and its projection onto the CHIPX axis, respectively. The data were binned with 30~counts per bin. The best-fitting {\tt wabs} $\times$ [BB$_{\rm low}+$ BB$_{\rm high}+$ PL] model is shown by solid histogram. We used the two BB components (BB$_{\rm low}+$ BB$_{\rm high}$) to simulate the `background' contribution from the wings of the pulsar's PSF (see text for details). 
 \emph{Right}: The 68\%, 90\% and 99\% confidence contours in the $\Gamma$--PL normalization plane for the extended emission for the simultaneous fit of both ACIS-S TE and CC-mode spectra. 
 The PL normalization is in units of $10^{-7}$ photons keV$^{-1}$ cm$^{-2}$. The dashed lines
are the lines of constant unabsorbed flux in the 0.5--8~keV energy range in units of $10^{-15}$ erg s$^{-1}$ cm$^{-2}$. \label{cont_ext}}
\end{figure*}

We assumed that the spectrum of the extended emission can be described by a PL model, as expected for a PWN. To determine the PL model parameters,
we fit the model {\tt wabs} $\times$ [BB$_{\rm low}+$ BB$_{\rm high}+$ PL$_{\rm ext}$] simultaneously to the two spectra described above (the total spectrum in the 3\farcs4--14\farcs8 annulus of the TE data, and the total spectrum in the rectangular region of the CC data). The two BB components are included to model the background pulsar emission and they are fixed to the values found from the regions extracted of the MARX simulation. The column density of the {\tt wabs} component was fixed to $N_{\rm H} = 4.3 \times 10^{20}$ cm$^{-2}$, the best-fit value found from the fit to the CC-mode spectrum (see Section \ref{pulsar_mod}). 
Table \ref{ext_fluxes} shows the resulting values of the photon index,
normalization and flux. The photon index is $\Gamma=7.7^{+1.6}_{-1.2}$ (the uncertainties are at the 90\% confidence level). The unabsorbed fluxes between 0.5--8 keV were computed by adding a CFLUX component to the above models. The data were group to 30 counts per bin. However, as the value of the $\chi^2$ statistic indicates (see Table \ref{ext_fluxes}), the fits to the extended emission are only marginally acceptable. As a result, the errors in the above fits are likely underestimated.
The resulting fit is shown in Figure \ref{cont_ext} (left panel), and the confidence contours for $\Gamma$ and the corresponding normalization are shown in Figure \ref{cont_ext} (right panel).

\begin{deluxetable}{lcccc}
\tablewidth{0pt}
\tabletypesize{\scriptsize}
\tablecolumns{5}
\tablecaption{Fit of the extended emission spectrum \label{ext_fluxes}}
\tablehead{ \colhead{Mode\tablenotemark{a}}&\colhead{$\Gamma_{\rm{ext}}$\tablenotemark{b}}&\colhead{$\mathcal{N}^{\rm ext}_{-7}$\tablenotemark{c}}&\colhead{$\chi_{\nu}^{2}$/dof}&\colhead{$F^{\rm un}_{-15}$\tablenotemark{e}}}
\startdata
TE+CC &
$7.7^{+1.6}_{-1.2}$
 & $6.09^{+11.2}_{-6.0}$  & 1.43/37 &
$8.3^{+5.7}_{-4.4}$
\enddata
 \tablecomments{The fits from {\tt wabs} $\times$ [BB$_{\rm low}+$ BB$_{\rm high}+$ PL$_{\rm ext}$] in the 0.3--8 keV energy range,  with
fixed $N_{\rm H}=4.3 \times 10^{20}$~cm$^{-2}$ (the best-fit value found from the fit to the CC-mode spectrum) and the values for the
BB components of the model fixed at the values derived from the fit to the MARX simulation of the pulsar spectrum (see text for details).}
\tablenotetext{a}{The mode of the ACIS-S observation.}
\tablenotetext{b}{The photon index for the putative extended emission.}
\tablenotetext{c}{The normalizations in units of $10^{-7}$~photons~keV$^{-1}$~cm$^{-2}$~s$^{-1}$ at 1 keV.}
\tablenotetext{d}{The value of $\chi^2$ statistic for binned data with minimum 30 counts per bin}
\tablenotetext{e}{Total unabsorbed flux for the extended emission in the 0.5--8 keV energy range, in units of $10^{-15}$ erg s$^{-1}$ cm$^{-2}$. }
\end{deluxetable}

\section{Discussion}
In this section, we discuss the putative extended emission with two possible interpretations: a PWN and a dust scattering halo.

\subsection{The PWN Interpretation of the Excess Emission\label{pwn_int}}

The detected excess of counts in the 3\farcs5--15\arcsec\ annulus around the pulsar,
with respect to a ChaRT/MARX point-source simulation, might be a PWN of PSR B0656+14. For $d=288$ pc, the unabsorbed luminosity of the putative PWN is $L_{\rm ext} =8.2_{-2.3}^{+3.7} \times 10^{28}$~erg~s$^{-1}$ in the 0.5--8 keV band, based on the flux derived from the TE-mode data for the PL model
(see Table \ref{ext_fluxes}). Although the corresponding radiative efficiency, $\eta_{\rm{pwn}}=L_{\rm ext}/\dot{E} \sim 2.2\times 10^{-6}$,
and the ratio $L_{\rm ext}/L_{\rm psr}^{\rm nonth}\sim 0.05$,
are lower than for majority of known PWNe \citep{karg08}, they are close to those for the Geminga pulsar \citep{pavl10},
which has a similar spin-down power and a factor of 3 larger characteristic age.
However, the approximately round (perhaps ring-like)
shape, as we might assume based on the lack of the statistically significant asymmetry (see Section \ref{asymm_sec}), and the extremely soft spectrum
($\Gamma_{\rm ext}\sim 8$) of the extended emission are very different from those of any other known PWN
(e.g., the Geminga PWN shows a bow-shock structure with a tail behind the
pulsar, and its spectrum is hard, $\Gamma_{\rm pwn}\sim 1$).

The PWN morphology depends on pulsar's Mach number ${\mathcal M}=v/c_s$, where $v$ is the pulsar's velocity, and $c_s$ is the speed of sound
in the ambient medium. Since even the transverse velocity of B0656+14, $v_\perp = v \sin i = 60^{+7}_{-6}$~km~s$^{-1}$  \citep{bris03},
where $i$ is the angle between the velocity vector and the line of sight, exceeds the typical speed of sound in the interstellar medium (ISM),
$c_{s} \sim 10$ --30 km s$^{-1}$, the pulsar moves supersonically, and the interaction of its wind with the ISM should produce
a bow-shock PWN. X-ray emission from such a PWN is due to synchrotron radiation from the shocked pulsar wind outside the termination shock (TS). For large ${\mathcal M}$  and small values of the magnetization parameter $\sigma$ of the preshock isotropic wind \citep{kenn84b}, the TS acquires a bullet-like shape \citep{bucc05} with a distance $R_h \simeq (\dot{E}/4\pi c p_{\rm ram})^{1/2}$ between the pulsar and bullet head,
where $p_{\rm ram}=\rho_{\rm amb} v^{2} = 1.67\times 10^{-10} n v_7^2$~dyn~cm$^{-2}$ is the ram pressure, $\rho_{\rm amb}$ is the
mass density of the ambient medium, $n = \rho_{\rm amb}/m_H$ (in units of cm$^{-3}$), and $v_7=v/(10^7\, {\rm cm/s})$.
The bullet's cylindrical radius is $r_{\rm TS}\sim R_h$ and the distance of its back surface from the pulsar is $R_b\sim 6 R_h$. The shocked wind outside the TS is confined by the contact discontinuity (CD) surface, which has an approximately cylindrical shape behind the bullet, with a radius
$r_{\rm CD} \sim 4 R_h$. The shocked wind flows along and past the bullet, reaching very high flow velocities,
$\sim (0.1$--$0.3)c$ in the ``inner channel'' ($r\lesssim r_{\rm TS}$) and up to $(0.8$--$0.9)c$ in the ``outer channel''
($r_{\rm TS} \lesssim r \lesssim r_{\rm CD}$) of the collimated outflow.
The appearance of such a PWN depends on the velocity orientation angle $i$. For $i$ close to $90^\circ$, the PWN would
look like a bow-shock structure accompanied by a tail of collimated wind outflow. However, for small $\sin i$
the PWN would look nearly round because of the projection effect. Since the emission from the nearly relativistic flow is predominantly directed along the flow velocity, the apparent PWN radius $R_{\rm pwn}$ is smaller, $R_{\rm pwn}\lesssim R_h$, in the case of approaching pulsar than in the case of receding pulsar, when $R_{\rm pwn}\gtrsim r_{\rm CD}\sim 4 R_h$.

For the parameters of PSR B0656+14, the bow-shock stand-off distance is
\begin{equation}
R_h \sim
0.013  n^{-1/2} \sin i\,\,
{\rm pc,}\,
\end{equation}
for $v_\perp = 60$ km s$^{-1}$ \citep{bris03}, which corresponds to the projected angular distance of
$\sim 10'' n^{-1/2} \sin^2i$ between the pulsar and the bullet head. The value of the
velocity orientation angle is unknown, but the lack of significant azimuthal asymmetry in the extended emission (see Section 3),
as well as the fact that $v_\perp$ is considerably smaller than the typical pulsar speed, $v\sim 400$--500 km s$^{-1}$, argue for
a small value of $\sin i$. The observed size of the extended emission can be reconciled with a small
$\sin i$ (hence small $R_h$), assuming the pulsar is moving from the observer. For instance, for $\sin i=0.15$ (i.e., $v=400$ km s$^{-1}$),
we obtain the (angular) radius $R_{\rm pwn}\sim 6'' n^{-1/2}$, which is close to the maximum radius of the extended emission,
$\sim 20''$, at $n\sim 0.1$ cm$^{-3}$. The annular appearance of the extended emission could be interpreted assuming that it
comes from the outer channel of the collimated tail outflow, $r_{\rm TS} \lesssim r \lesssim r_{\rm CD}$, while the emission from the inner channel, $r\lesssim r_{\rm TS}$, is too faint to distinguish it from the wings of the pulsar's PSF.

Note that there is substantial evidence that the pulsar spin axis is often aligned with the velocity vector \citep{roma09}. If this is the case for
PSR B0656+14, the small velocity orientation angle $i$ implies a small angle $\zeta$ between the pulsar spin axis and the line of sight.
Small values of the spin orientation angle, $\zeta = 9^\circ\pm 1^\circ,$ were found by \citet{pier14} from fitting the $\gamma$-ray light curve
of PSR B0656+14 with the so-called Polar Cap (PC) gap model \citep{musl03} which gave a better fit than other models (such as the Outer Gap
model), in contrast to the majority of $\gamma$-ray pulsars. The same fit yielded the magnetic obliquity (the angle between the spin
and magnetic axes) $\alpha = 10^\circ\pm 1^\circ$, which classifies PSR B0656+14 as a nearly aligned rotator \citep[see also][]{ever01,lloy03}.

Thus, the observed size and shape of the putative PWN could be explained by the small angle between the line of sight and
the velocity of the receding pulsar. It is more difficult to explain the softness of the extended emission spectrum,
$\Gamma_{\rm ext} \approx$ 8  (see Table \ref{ext_fluxes}), which is much softer than the spectra of other PWNe, where the photon indices
are in the range of $\sim 1$--3 \citep{karg08}. We can only speculate that the unusually soft PWN spectrum might be
associated with peculiar $\gamma$-ray properties of PSR B0656+14, such as a very low $\gamma$-ray efficiency, $\eta_\gamma =0.0084\pm 0.0008$,
unusually soft spectrum, $\Gamma_\gamma = 2.4\pm 0.4$, and low cutoff energy, $E_{\rm cut}=0.7\pm 0.5$~GeV, which all might be caused by smallness of the angle between the magnetic and spin axes.

\subsection{Is the Extended Emission a Dust Scattering Halo?\label{pwn_dust}}

Since the image of any X-ray source is surrounded by a halo due to interstellar dust scattering \citep{over65},
such a halo could contribute to the extended emission seen in PSR B0656+14. For an optically thin dust, the halo spectral intensity
(in units of, e.g., photons~cm$^{-2}$~s$^{-1}$~keV$^{-1}$~arcmin$^{-2}$) at photon energy $E$ and angle $\theta$ from the central source,
and the halo spectral flux in an annulus $\theta_1<\theta<\theta_2$ (photons cm$^{-2}$ s$^{-1}$ keV$^{-1}$), are
\begin{align}
I_{\rm{halo}}(\theta,E)=F_{\rm{c}}(E)N_{\rm{H}}\int_0^1{dx \frac{f(x)}{x^{2}}\frac{d\sigma_{\rm{s}}(E,\theta_{\rm{s}})}{d\Omega_{\rm{s}}}}\label{eqSB}
\end{align}
and
\begin{align}
F_{\rm halo}(\theta_1,\theta_2;E)=2\pi\int_{\theta_1}^{\theta_2} I_{\rm halo}(\theta, E) \theta\, d\theta\,.
\end{align}
Here $F_{\rm{c}}(E)$ is the spectral flux of the central point source, $N_{\rm{H}}$ is the interstellar hydrogen column density,
$x$ is the ratio of the distance between the source and the scatterer to the distance between the source and the observer,
$f(x)$ is the dust density distribution along the line of sight ($\int_0^1 f(x)\,dx =1$),
 $d\sigma_{\rm{s}}/d\Omega_{\rm{s}}$ is the differential scattering cross section per one hydrogen atom, averaged over the dust grain distribution over sizes and other grain properties, and $\theta_s \simeq \theta/x$ is the scattering angle (see, e.g., \citeauthor{math91} \citeyear{math91}).

In the Rayleigh-Gans approximation \citep{mauc86,math91}, which is applicable at energies $E\gtrsim 0.5$ keV,
at least for small grains \citep{smit98}, \cite{drai03b} suggested a convenient analytical form for the
averaged differential cross section:
\begin{equation}
\frac{d\sigma_{\rm{s}}(E,\theta_{\rm{s}})}{d\Omega{\rm_{s}}} \approx \frac{\sigma_{\rm{s}}}{\pi \theta_{\rm{s,50}}^{2}}\frac{1}{(1+\theta_{\rm{s}}^{2}/\theta_{\rm{s,50}}^{2})^{2}}\,,
\end{equation}
where
  \begin{equation}
  \theta_{\rm{s,50}} \approx \frac{\Theta}{E} \quad {\rm and}\quad
  \sigma_{\rm{s}} \approx \frac{S}{E^{2}} 10^{-22} \rm{cm}^{-2}
\label{eqthetas,50}
  \end{equation}
are the median scattering angle and the total cross section, respectively. In the above equations the factor $S$ is a constant of the order of 1
(e.g., $S\simeq0.5$ according to \citeauthor{pred95} \citeyear{pred95}), $E$ is the photon energy in keV, and the constant $\Theta$ depends on the dust model (e.g., \citeauthor{drai03b} \citeyear{drai03b} derived $\Theta=360 \arcsec$ for the model by \citeauthor{wein01} \citeyear{wein01}; see \citeauthor{dura11a} \citeyear{dura11a} for more discussions). Note that in this model the optical depth of the dust between the
source and the observer  model is
\begin{equation}
\tau_{\rm sca}(E) = N_{\rm H}\sigma_s = N_{\rm H,22} S E^{-2}\,,
\label{opt_depth}
\end{equation}
where $N_{\rm H,22}=N_{\rm H}/(10^{22}\,{\rm cm}^{-2})$.

 In the case of uniform dust distribution along the line of sight, $f(x)=1$, the above equations give
 \begin{align}
  I_{\rm{halo}}(\theta,E)=F_{\rm{c}}(E)N_{\rm{H,22}}\frac{S}{2\pi\Theta\theta E}\notag\\ \times \left[\arctan\frac{\Theta}{\theta E}-\frac{(\Theta/\theta E)}{1+(\Theta/\theta E/)^{2}}\right]\,,
\label{halo_int_uniform}
 \end{align}
\begin{align}
F_{\rm halo}(\theta_1,\theta_2;E)=F_c(E) N_{\rm H,22}\frac{S}{\Theta E}
\notag\\
\times \left(\theta_2\arctan\frac{\Theta}{\theta_2E}-\theta_1\arctan\frac{\Theta}{\theta_1E}\right)\,.
\label{halo_flux_uniform}
\end{align}
For angles $\theta$ much smaller than the characteristic halo size,
$\theta_h\sim \Theta/E$, Equations (\ref{halo_int_uniform}) and (\ref{halo_flux_uniform}) turn into
\begin{equation}
I_{\rm{halo}}(\theta,E)\approx F_{\rm{c}}(E) N_{\rm{H,22}}\frac{ S}{4\Theta\theta E}
\label{halo_int_unif_limit}
\end{equation}
and
\begin{equation}
F_{\rm halo}(\theta_1,\theta_2;E)\approx
F_c(E) N_{\rm H,22}\frac{\pi S}{2\Theta E} \left(\theta_2 - \theta_1\right)\,.
\label{halo_flux_unif_limit}
\end{equation}

For comparison with observations, the halo spectral intensity and flux should be convolved with the detector's response. If the halo
size is much larger than the PSF size, and the energy redistribution in the detector is not important, the observed surface brightness
(counts~s$^{-1}$~arcmin$^{-2}$) and count rate in an annulus (counts~s$^{-1}$) can be estimated by convolving the spectral intensity and flux with
the detector's effective area $A_{\rm eff}(E)$ \citep[see, e.g.,][]{dura11a}. If the size of detected extended emission
is much smaller than the expected halo size (like in the case of PSR B0656+14), we can use Equations (\ref{halo_int_unif_limit})
and  (\ref{halo_flux_unif_limit}) to obtain the annulus count rate
\begin{equation}
C_{\rm halo}(\theta_1,\theta_2) \approx
C_c N_{\rm H,22}S\langle E^{-1}\rangle
\frac {\pi (\theta_2-\theta_1)}{2\Theta}
\label{halo_crate_unif}
\end{equation}
and the normalized halo profile (the ratio of the surface brightness to the central source count rate, in units of arcmin$^{-2}$)
\begin{equation}
J_{\rm halo}(\theta)
\approx \frac{N_{\rm H,22} S
\langle E^{-1}\rangle}{4\Theta\theta}\,,
\label{halo_profile_unif}
\end{equation}
where
$C_c=\int F_{\rm c}(E) A_{\rm eff}(E)\,dE$ is the observed count rate of the central source (in units of counts s$^{-1}$) and
\begin{equation}
\langle E^{-1}\rangle =
\left(
C_c\right)^{-1} \int F_{\rm c}(E) A_{\rm eff}(E) E^{-1}\,dE\,
\end{equation}
is the mean inverse energy.

We used Equations (\ref{halo_crate_unif}) and (\ref{halo_profile_unif}) and the HRC-I data to check if the observed extended emission around PSR B0656+14 can be interpreted as a dust scattering halo. We used the HRC-I observation because it provided more counts than the very short ACIS-S TE mode observation, and it did  not suffer from pileup.

Using the HRC-I effective area\footnote{http://cxc.harvard.edu/proposer/POG/html/chap4.html} and the pulsar spectral model given in Table 3,
we estimated $\langle {E}^{-1} \rangle=1.48$ (keV$^{-1}$). Assuming that the extended emission in the 3\farcs4--14\farcs8 annulus is due to dust scattering, we used the observed excess count rate in the annulus,
$C_{\rm halo}(3\farcs4,14\farcs8) =0.0105\pm  0.0011$~counts~s$^{-1}$ (see Table 1), and  the pulsar count rate
$C_c=1.3070\pm 0.0058$~counts~s$^{-1}$, to obtain the scattering optical depth at 1~keV,
\begin{equation}
\tau_{\rm sca}(1\,{\rm keV}) = N_{\rm H,22} S = (0.109\pm  0.011) (\Theta/360'').
\label{tausca1}
\end{equation}
For $N_{\rm H,22}=0.043\pm 0.007$, determined from the fit of the pulsar spectrum, we obtain $S=(2.53\pm 0.68)(\Theta/360'')$. This value substantially exceeds the typical value $S\approx 0.5$ \citep{pred95}. Although the values of $S$ measured for different sources show a strong
scatter, likely due to different dust properties and gas-to-dust ratio, our estimate for PSR B0656+14 is much larger than the maximum value $S\approx0.55$ for the sources investigated by \cite{pred95}. Moreover, based on the tight correlation $\tau_{\rm sca}(1\,{\rm keV})=(0.056\pm 0.01) A_V$ found by \cite{pred95}, our estimate corresponds, for $\Theta=360''$, to the optical extinction $A_V=1.94\pm 0.40$ (or color index $E_{B-V}\approx 0.63\pm 0.13$), too large for PSR B0656+14 \citep{dura11}. Therefore, it seems unlikely that the detected extended emission is solely due to dust scattering.

\begin{figure}
\includegraphics[scale=0.85,angle=0]{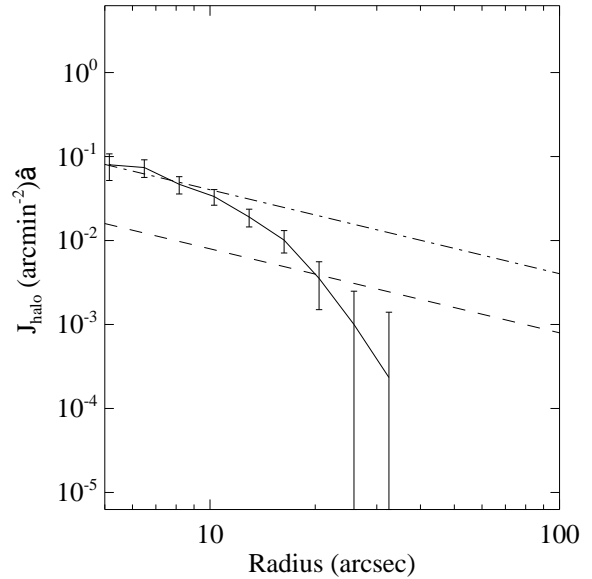}
\caption{The observed normalized profile in the HRC-I image of the  extended emission
around PSR B0656+14 (errorbars) and the model dust
scattering halo profiles for $\tau_{\rm sca}(1\,{\rm keV}) = 0.109$ and 0.021
(dash-dot and dashed lines, respectively).
\label{halo_profile}}
\end{figure}

An additional argument against the scattering halo interpretation is provided by the comparison of the normalized surface brightness profiles of the detected emission and the halo model. For the $\tau_{\rm sca}(1\,{\rm keV})$ given by Equation (\ref{tausca1}), we obtain
$J_{\rm halo}(\theta) = (0.403\pm 0.041)\,\theta^{-1}\,\, {\rm arcmin}^{-2}$, where $\theta$ is in arcseconds.
This model profile is shown in Figure \ref{halo_profile} (dash-dotted line) together with the normalized profile of the excess emission detected with HRC-I (to calculate the excess we subtracted the PSF model from the MARX simulation with {\em ditherblur} = 0\farcs25 as it gave the best fit to the pulsar's PSF at $\theta\lesssim 3''$ -- see Figure 3). We see from Figure \ref{halo_profile} that the model profile decreases with $\theta$ much slower than the measured profile and
exceeds the latter at $\theta\gtrsim 10''$. If, instead, we assume $\tau_{\rm sca}(1\,{\rm keV}) = 0.021$, which corresponds to $S=0.5$ at $N_{\rm H,22}=0.043$ and $\Theta=360''$, the normalized model profile,
$J_{\rm halo}(\theta) = 0.0795\,  \theta^{-1}\,\, {\rm arcmin}^{-2}$,
shown by the lower dashed line in Figure \ref{halo_profile},
goes substantially below the observed one at $\theta \lesssim 15''$.
It is clear from these examples that the model halo profile would not fit
the observed profile at any value of $S$, or $\tau_{\rm sca}(1\,{\rm keV}$),
at least for the uniform dust distribution. The main reason for that is that the size of the
detected extended emission is much smaller than the dust halo size, $\theta_h\sim 9'$.

Thus, it is unlikely that the extended emission is due entirely to dust scattering. However, we cannot exclude a possibility
that the extended emission is a combination of both a PWN and a dust scattering halo. In this case, the total spectrum of the extended emission, shown in Figure 6, would be a combination of the very soft halo spectrum (which is even softer than the pulsar's spectrum)
and a (presumably hard) PWN spectrum, and a significant halo contribution could explain the softness of the extended emission
spectrum. Figure 8 suggests that the PWN and halo contributions would dominate at smaller and larger angular distances from the
pulsar, respectively, i.e., the spectrum of the extended emission would depend on $\theta$.
To check this hypothesis, deep observations of the extended emission should be carried out with sufficiently high  angular and
spectral resolutions to provide enough counts for spatially resolved spectroscopy of the extended emission.
This can only be done with the {\sl Chandra} ACIS detector in imaging (TE) mode, with an exposure much longer than the 5 ks exposure of ACIS-S TE mode
observation used in this work. To disentangle the PWN and halo contributions, the spectra in several annuli
around the pulsar should be compared with model halo + PWN spectra, taking into account the energy dependence of the halo profile.

\section{Summary}

Using \emph{Chandra} observations, we looked for extended emission around PSR B0656+14, by performing point source simulations using ChaRT and MARX. We compared the observed surface brightness with the simulated pulsar's surface brightness, and we found evidence of faint extended emission in a 3\farcs5--15\arcsec\ annulus around the pulsar,
with a luminosity $L_{\rm ext}\sim 8 \times 10^{28}$ erg s$^{-1}$ in the 0.5--8 keV band and a soft spectrum. This emission could be
a combination of a PWN and a dust scattering halo, dominating at smaller and larger distances from the pulsar, respectively.
Although we expect a bow-shock PWN for the supersonically moving pulsar, the observed size and shape of the
extended emission, small and apparently round, is consistent with the PWN interpretation
for its inner part if the pulsar's velocity has a large radial component directed away from the observer.
The soft spectrum of the extended emission could be due to the contribution of the dust scattering halo.
With data obtained from a deeper ACIS observation in TE mode, this contribution could be modeled as an additional background to the PWN emission, and the surface brightness distribution and the spectrum of the PWN component could be inferred.

\acknowledgements
\small{Support for this work was provided by the National Aeronautics and Space
Administration through {\sl Chandra} Award Number GO3-4095 issued by the
{\sl Chandra X-ray Observatory}  Center, which is operated by the Smithsonian Astrophysical Observatory for and on behalf of the National Aeronautics Space Administration under contract NAS8-03060. The work was also partly supported by the NASA grant NNX09AC84G.
L.B. thanks David Rafferty for carefully reading the draft and his help with the X-ray data reduction, Mike Wise for helping with MARX, Paul Nulsen for carefully reading a earlier version of the manuscript and useful discussions about the {\sl Chandra} PSF, and Elisa Costantini for some discussions about the dust halo scenario. Also, we thank the referee for the detailed comments and suggestions which further improve and clarify the draft.}

\medskip\noindent
{\em Facility:} \facility{{\sl CXO}}

 \appendix
\section{Comparison to known {\sl Chandra} calibrators}

In order to check whether the point source simulations systematically underestimate the observed surface brightness in the
3\arcsec -- 20\arcsec\ annulus, we performed the same analysis as in Section \ref{S:profiles} for  two objects listed as calibrators for the {\sl Chandra} PSF\footnote{see \url{http://www.cxc.harvard.edu/cal/docs/cal\_present\_status.html}} -- the quasar 3C273 and the binary star system AR Lac.
These sources have very low absorbing columns, $N_{\rm H} =1.8 \times 10^{18}$~cm$^{-2}$ for AR Lac  \citep{huen03}  and $N_{\rm H} =1.79 \times 10^{20}$~cm$^{-2}$ for 3C273 \citep[galactic absorption value derived from ][]{dick90}, hence
negligibly faint dust scattering halos.

\begin{figure*}
\plottwo{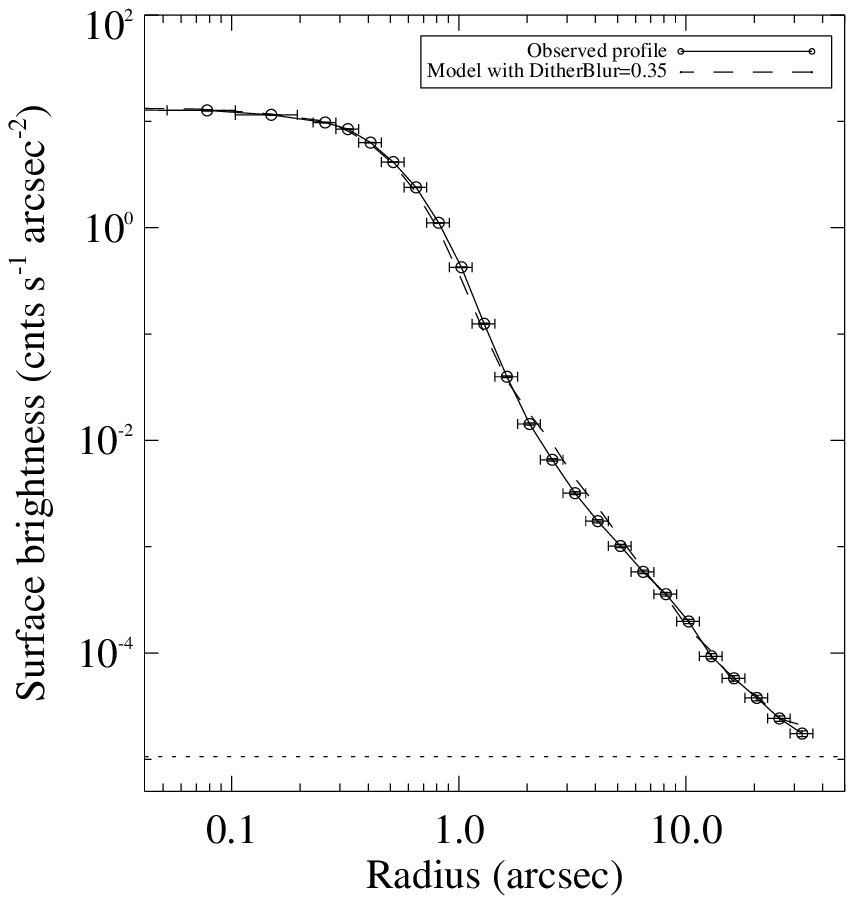}{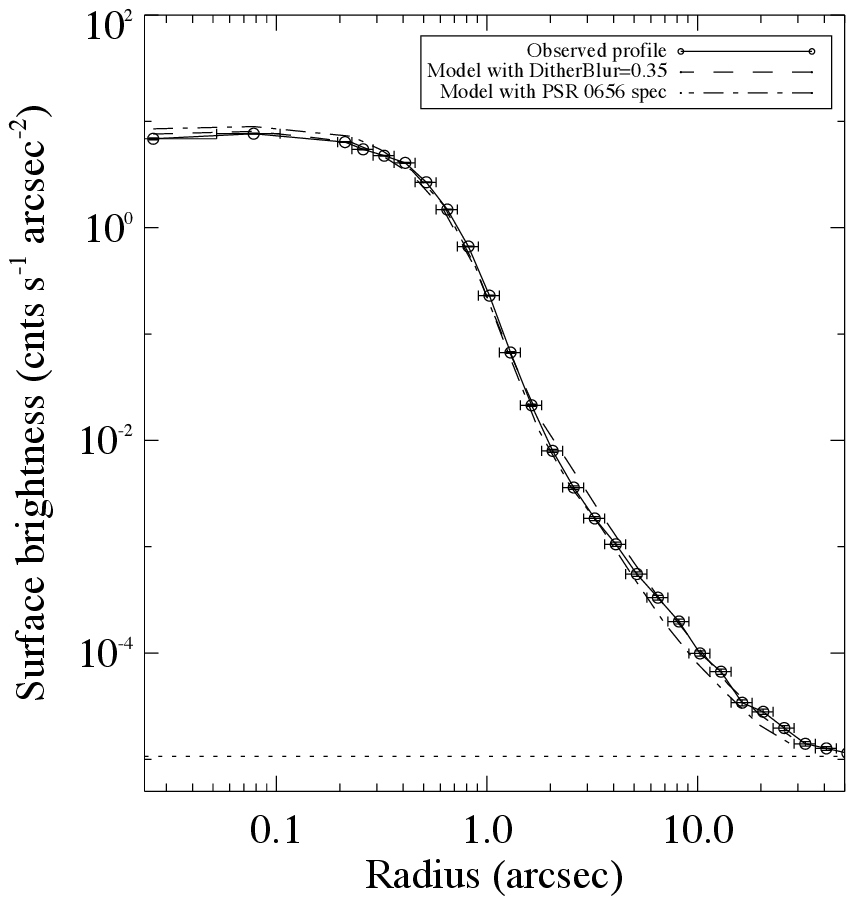}
\caption{Surface-brightness profiles for the HRC-I  observation of 3C273 (symbols and error bars) in the left panel and for the HRC-I observation of AR Lac (symbols and error bars) in the right panel.  The simulations  are for \textit{ditherblur}$=0.35$\arcsec. \label{F:3c273_ARLac}}
\end{figure*}

For 3C273, we reduced the HRC-I observation of 2000 January 22 (ObsID 461) and the ACIS-S observation of 2000 June 20 (ObsID 1711).
Because of the pileup contamination for the inner parts of the ACIS-S PSF, we compare the surface brightness profiles between the observation and simulation for the HRC-I instrument. We used the ACIS-S observation to create an input spectrum for the ChaRT and MARX simulators.
To exclude the region affected by pileup, we extracted the spectrum from the 5\arcsec--10\arcsec\ annulus after masking the
emission from the 3C273's jet. The spectrum from this region may differ somewhat from a spectrum that includes the full point-source emission, but non-piled spectral data for the central region are not available.
The spectrum was modeled as a {\tt wabs} $\times$ PL with
$N_{\rm H} = 1.79 \times 10^{20}$ cm$^{-2}$,
 $\Gamma=1.54 \pm 0.03$ and ${\cal N} =2.02 \pm 0.04\times 10^{-4}$ photons keV$^{-1}$cm$^{-2}$s$^{-1}$ at 1 keV. The 200~ks simulated profile and the observed profile are compared in Figure \ref{F:3c273_ARLac} (left panel). It is clear from this figure that the simulated PSF is a good match to the observed one at all radii, except for a slight excess in the simulated profile between 2\arcsec\ and 4\arcsec.

For AR Lac, we reduced the HRC-I observation from 2000 September 11 (ObsID 1385) and the HETG ACIS-S observation from 2000 September 11 (ObsID 6).
We extracted first-order spectra from the HETG data and fit them with a MEKAL model in the Sherpa package. The resulting best-fit spectrum ($kT=0.986$ keV, $Z=0.128$ $Z_{\odot}$, ${\cal N} =0.048$) was then used as input to ChaRT/MARX to simulate a 200 ks HRC-I observation. We compare the simulated and observed HRC-I profiles in Figure \ref{F:3c273_ARLac} (right panel). Again, there is a good match between the two profiles, with a slight excess in the simulated profile between 2\arcsec--4\arcsec\  (similar to 3C273, see Figure \ref{F:3c273_ARLac} left panel). In neither the AR Lac profile nor the 3C273 profile do we see evidence for excess extended emission relative to the simulated profiles such as that seen in the HRC-I observation of PSR B0656+14.

\begin{figure*}
\plottwo{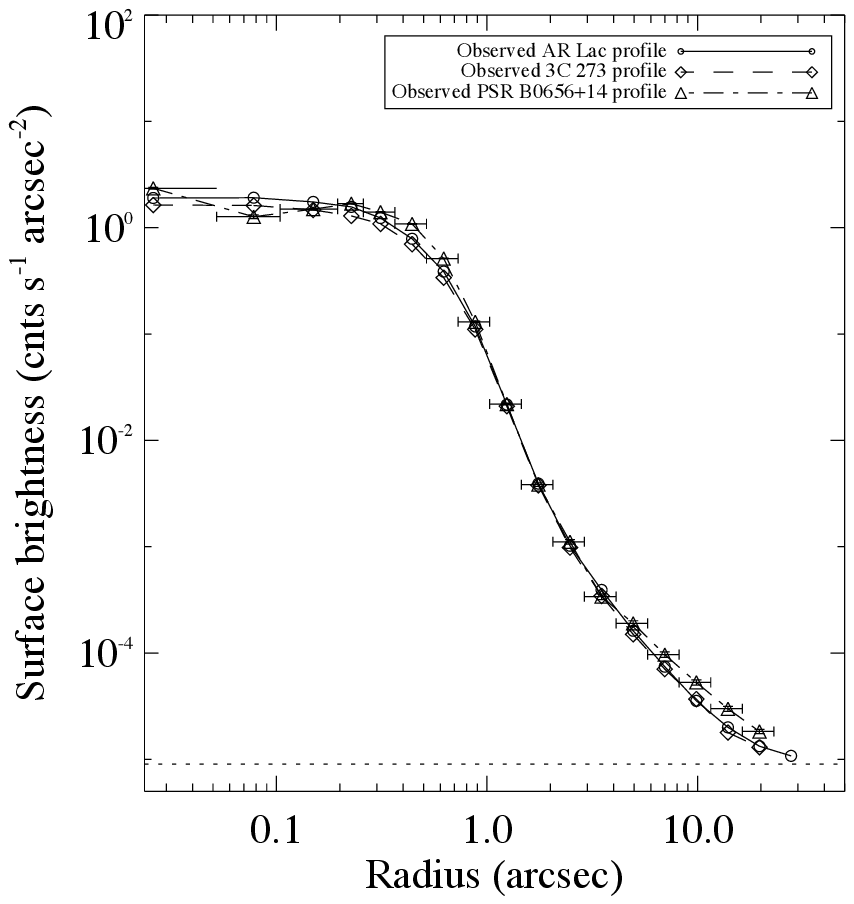}{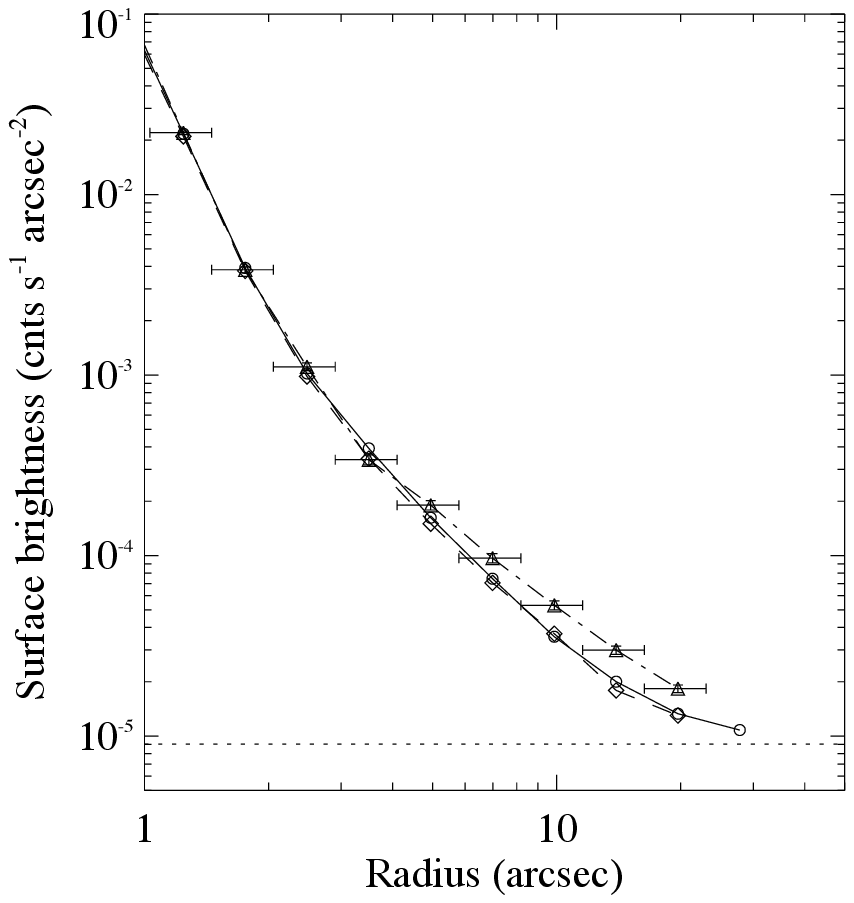}
\caption{\emph{Left:} Comparison of the surface-brightness profiles for the HRC-I observations of PSR~B0656+14 (triangles), 3C273 (diamonds) and AR Lac (circles). \emph{Right:} The same as the left panel, but zoomed in on the region that encompasses the putative extended emission (3\arcsec $< r < 20$\arcsec).\label{F:all_obs}}
\end{figure*}

Lastly, we compare the observed HRC-I profiles for PSR B0656+14, 3C273, and AR Lac in Figure \ref{F:all_obs}. In this plot, the three profiles have been corrected for exposure with \emph{mkexpmap} in CIAO using the spectra extracted above for weights. The resulting profiles were then normalized to the surface brightness of PSR B0656+14 at $r=1$\arcsec. For the comparison, local backgrounds of $1.06\times 10^{-4}$ and  $1.07\times 10^{-5}$ cnts s$^{-1}$ arcsec$^{-2}$ were subtracted from the 3C273 and AR Lac profiles, respectively, and the local background of
$0.903\times 10^{-5}$ cnts s$^{-1}$  arcsec$^{-2}$ for PSR B0656+14 was added so that the backgrounds in all three observations are the same. The profile of PSR B0656+14 is slightly more peaked at the center and shows an excess at radii beyond $r\approx 3''$--4\arcsec. While this excess is not great, it is systematic and should be noted that the softer spectrum of PSR B0656+14 should result in a more centrally concentrated profile than the two comparison profiles, not the less centrally concentrated one seen in Figure \ref{F:all_obs}.

\bibliographystyle{apj}
\bibliography{b0656pwn_final_resubmit.bbl}

\end{document}